\documentclass[aps,prl,twocolumn,showpacs,superscriptaddress,floatfix,footinbib,10pt]{revtex4-1}
\usepackage{amssymb}
\usepackage{graphicx}
\usepackage{amsmath}
\usepackage{amsbsy}
\usepackage{amsthm}
\usepackage{bbm}
\usepackage{bm}
\usepackage{epsfig}
\usepackage{dsfont}
\usepackage{hyperref}
\usepackage{soul}
\usepackage[usenames,dvipsnames]{xcolor}
\DeclareMathOperator{\atanh}{atanh}

\begin{document}


\title{Relativistic motion generates quantum gates and entanglement resonances}
\date{\today}

\author{David Edward Bruschi}
\affiliation{School of Mathematical Sciences, University of Nottingham, University Park,
Nottingham NG7 2RD, United Kingdom}
\affiliation{School of Electronic and Electrical Engineering, University of Leeds, Leeds LS29JT, United Kingdom}

\author{Andrzej Dragan}
\affiliation{School of Mathematical Sciences, University of Nottingham, University Park,
Nottingham NG7 2RD, United Kingdom}
\affiliation{Institute of Theoretical Physics, University of Warsaw, Ho\.{z}a 69, 00-049 Warsaw, Poland}

\author{Antony R. Lee}
\author{Ivette Fuentes}\thanks{Previously known as Fuentes-Guridi and Fuentes-Schuller.}
\author{Jorma Louko}
\affiliation{School of Mathematical Sciences, University of Nottingham, University Park,
Nottingham NG7 2RD, United Kingdom}

\begin{abstract}
We show that the relativistic motion of a quantum system can be used to generate quantum gates.
The nonuniform acceleration of a cavity is used to generate well-known two-mode quantum gates 
in continuous variables. 
Observable amounts of entanglement between the cavity modes are produced through resonances 
which appear by repeating periodically any trajectory.
\end{abstract}

\maketitle

\emph{Introduction.\thinspace---} 
Understanding how motion and gravity affect quantum information is a key feature in the implementation of new relativistic quantum technologies, including quantum cryptography and teleportation, in relativistic regimes and space-based scenarios that are currently under investigation~\cite{bb:spacebased}.  Recent results in relativistic quantum information show that the non-uniform motion of a cavity creates entanglement between the cavity field modes \cite{friis-gener,NFriis:IFuentes:2012}. In this Letter we employ the moving cavity scenario to show that the relativistic motion of a quantum system can be used to implement quantum gates. The gates can be readily implemented experimentally thanks to cutting-edge technology in superconducting circuits where the relativistic motion of boundary conditions has been demonstrated \cite{wilsonnature}. 

Finding suitable ways to store and process information in a quantum and relativistic setting is a main goal in the field of relativistic quantum information \cite{alsing2012}. Moving cavities are good candidates to store information \cite{BruschiAlpha,friis2011,Downes} since confined fields can be realized experimentally \cite{raimond2001} and observers can directly access their states by means of local operations. When a cavity is accelerated for a finite time the cavity modes are affected by the motion. A ~mismatch between the vacua at different times gives rise to the creation of particles  \cite{birrell-davies} which populate and entangle the modes. 

\begin{figure}[b!]
\includegraphics[width=\linewidth]{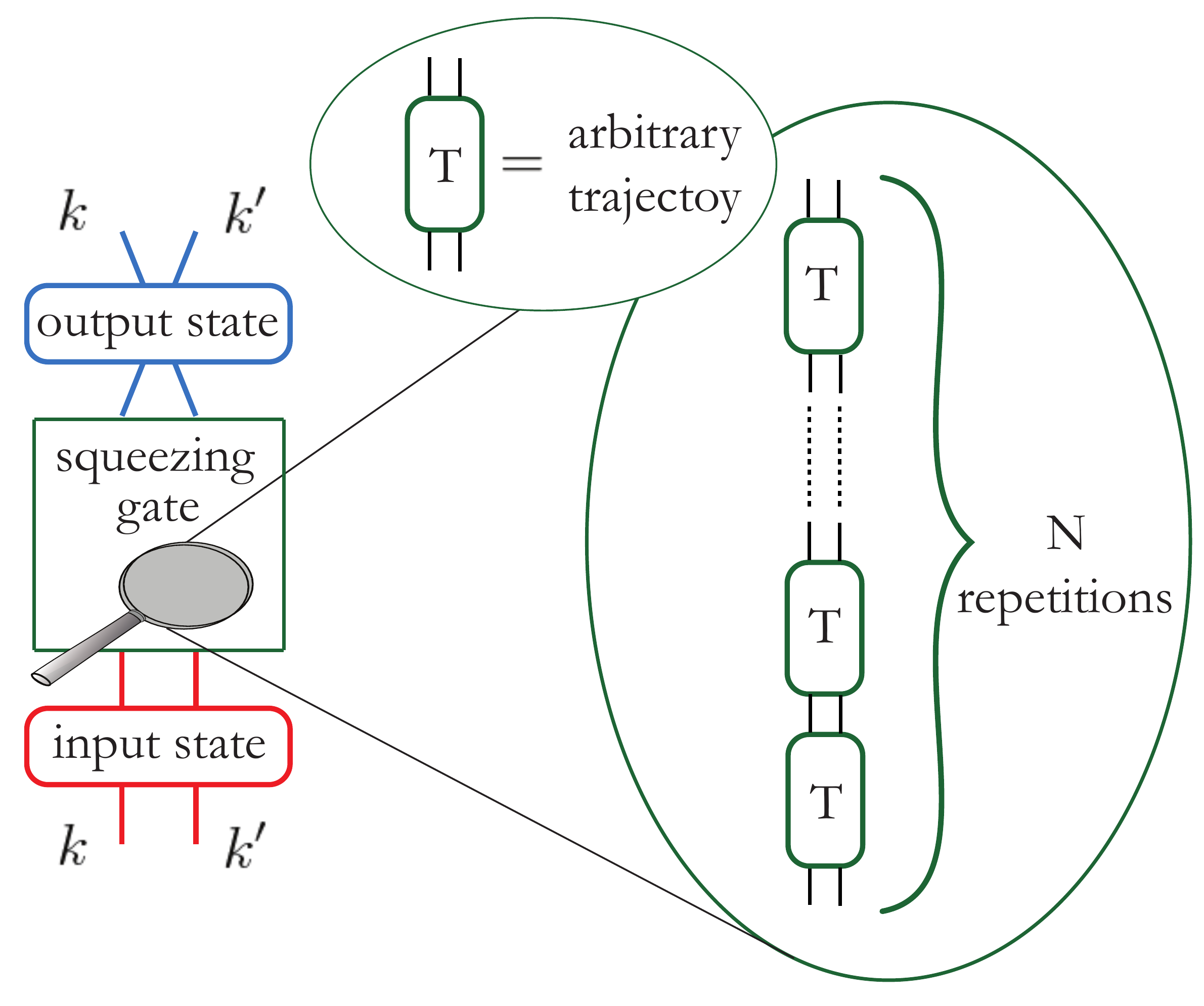}
\caption{\label{tmsgate2} The relativistic motion of the cavity is used to produce a two mode squeezing gate. The input ({\color{red}red}) and output ({\color{blue}blue}) states are Gaussian states. The non-uniform motion of the cavity produces two-mode squeezing. The procedure to enhance the effect is illustrated in the magnified {\color{Green}green} oval: by repeating $N$ times an arbitrary trajectory characterized by a total proper time $T$ (represented by the single box in the smaller oval at the top), the degree of squeezing is linearly increased.} 
\end{figure}

We show that two-mode squeezing gates, which are paradigmatic gates in continuous variable systems \cite{weedbrook2012}, can be produced when the cavity follows a trajectory which includes non-uniform acceleration. The amount of entanglement generated by the quantum gate can be enhanced through a resonance produced by repeating any trajectory periodically. 
The trajectory can be, for example, a return trip to Alpha Centauri or a short segment of uniform acceleration. We show analytically that for \textit{any\/} pair of oddly separated modes it is possible to find a travel time where the entanglement produced by the quantum gate increases linearly with the number of repetitions. These resonances appear independently of the details of the trajectory though the amount of entanglement generated does depend on the trajectory itself. Our scheme is illustrated in Fig.~\ref{tmsgate2}.

We present a class of sample travel scenarios in which the generated entanglement can be expressed analytically in terms of the magnitude and direction of the acceleration. This class includes the sinusoidal motion that is often considered in the dynamical Casimir effect literature~\cite{dodonov1990}. 

In brief, our main contribution is to implement quantum processing gates in in relativistic quantum field theory. 
This is a step beyond the various proposed nonrelativistic implementations of continuous variable quantum gates~\cite{yonezawa2010,weedbrook2012,Ukai2011}.

\emph{Setup.\thinspace---} 
We consider a real scalar field $\phi$ of mass $m$ contained within a cavity in $(1+1)$-dimensional spacetime. Boundary conditions are imposed such that the field vanishes at the cavity walls. The massless field can be treated as a special case of our study and the effect of additional transverse dimensions can be included as a positive contribution to the mass. The cavity follows a worldtube which is composed of segments of inertial and uniformly accelerated motion. We will start by describing the field within the cavity during such segments as seen by a comoving observer. 

During segments of inertial motion, we use Minkowski coordinates $(t,x)$ to describe our system. The cavity walls are placed at $x=x_{A}$ and $x=x_{B}$ where $0< x_A < x_B$, so that $L=x_B-x_A$ is the length of the cavity. The positive frequency mode functions with respect to the time translation Killing vector $\partial_t$ are:
\begin{align}
\label{MinkowskiSolutions}
\phi^{M}_k (t,x) &= \frac{1}{\sqrt{\omega_{k}L}}\sin\left[\frac{k\pi}{L}(x-x_{A})\right]e^{-i\omega_{k}t} \ ,
\end{align}
where $\omega_{k}= \sqrt{{(k\pi/L)}^2+m^2}$ are the Minkowski frequencies and $k\in\mathbb{N}$ (we set $c = \hbar =1$). 

We employ Rindler coordinates $(\tau,\chi)$ to describe the field during segments of uniformly accelerated motion. The transformations between Rindler and Minkowski coordinates are $\chi=\sqrt{x^{2}-t^{2}}$ and $\tau=(1/a)\atanh (t/x)$, where $a>0$. A uniformly accelerated observer comoving with the cavity follows the boost Killing vector field $\partial_{\eta}=x\partial_{t}+t\partial_{x}$ which in Rindler coordinates takes the form $\partial_{\eta}=(1/a)\partial_{\tau}$. The cavity walls are placed at $\chi=x_{A}$ and $\chi=x_{B}$ and the proper time and acceleration at the centre of the cavity are given by $\tau$ and $a=2/(x_{B}+x_{A})$, respectively. 

The solutions $\phi^{R}_k (\tau,\chi)$ to the Klein-Gordon equation that are of positive frequency with respect to $\partial_{\eta}$ can be expressed in terms of modified Bessel functions~\cite{nist-dig-library,Friis:Lee:Louko:13}. 
The Rindler frequencies are  $\Omega_k>0$ and are determined by $\phi^{R}_k (\tau,x_B) =0$. In the massless case the mode functions and the frequencies reduce to simple expressions~\cite{BruschiAlpha}. 

The quantized field operators are given by $\phi^M = \sum_k \bigl(\phi^{M}_k a_k + \text{h.c.} \bigr)$, and $\phi^R= \sum_k \bigl(\phi^{R}_k A_k + \text{h.c.} \bigr)$ during segments of inertial and accelerated motion, respectively. The Minkowski and Rindler annihilation and creation operators obey the standard commutation relations $\bigl[a_k, a^\dagger_l\bigr] = \delta_{kl}$ and  $\bigl[A_k, A^\dagger_l\bigr] = \delta_{kl}$.  

We will work in the covariance matrix formalism which is applicable to systems consisting of a discrete number of bosonic modes as long as the analysis is restricted to Gaussian states. In this framework the state of the system is entirely described by its first and second moments~ \cite{braunstein2005,adesso2005}. The evolution of the state is given by a similarity transformation $S^{Tp}\Delta S$ where $S$ is the symplectic representation of the evolution and $\Delta$ is the covariance matrix encoding all information pertinent to the state. $Tp$ denotes matrix transposition.

\emph{Travel scenario techniques.\thinspace---} 
Changes from inertial to accelerated motion and vice versa are implemented by the action of Bogolubov transformations.  Consider that at $t=0$ a cavity initially at rest begins to accelerate. The inertial and accelerated cavity modes are related by the Bogolubov transformation $\phi^{R}_k = \sum_n \, 
\bigl( \alpha_{kn} \phi^{M}_n + \beta_{kn} \phi^{M*}_n \bigr)$, where the star denotes complex conjugation
and the Bogolubov coefficients $\alpha_{kl}$ and $\beta_{kl}$ can be evaluated by taking Klein-Gordon inner products of the two sets of modes at $t=0$~\cite{birrell-davies,fabbri-navarro-salas,Crispino2008}. 

We employ a perturbation expansion of the Bogolubov coefficients such that 
$\alpha =\mathbb{I} + h\alpha^{(1)} + O(h^2)$ and $\beta = h\beta^{(1)} + O(h^2)$ 
where $h=aL$ is a small dimensionless expansion parameter~\cite{BruschiAlpha,friis2011}. 
For cavities of typical laboratory sizes, the small $h$ regime can accommodate extremely large accelerations. 
There are no restrictions on the duration, covered distance, or the achieved velocity of the motion. 


During the segments of inertial or accelerated motion the modes undergo free evolution which induces phase rotations on the state of the form $
U(\tau)=\bigoplus_{i=1}^{\infty} R(\theta_{i})$ 
where $R(\theta_i)$ is the standard $2\times2$ rotation matrix of angle $\theta_i$. The angles are given by $\theta_k=\omega_k t$ during coasting segments and $\theta_k=\Omega_k \tau$ during acceleration. The Rindler and Minkowski frequencies coincide to first order, 
$\Omega_k(h)=\omega_k+\mathcal{O}(h^2)$.  

We construct the cavity trajectories by composing these basic transformations. The field modes of a cavity initially at rest and the modes after any travel scenario are related through general  Bogolubov transformations.  In the covariance matrix formalism, transformations are represented the symplectic matrix $\mathcal{S}$, which can be decomposed into $2\times2$ blocks $s_{kk'}$ of the form
\begin{eqnarray}
s_{kk'}\label{nico}&=&
\left( \begin{array}{cccc}
\Re({A_{kk'}- B_{kk'}})&\Im({A_{kk'}+B_{kk'}})\\
-\Im({A_{kk'}- B_{kk'}})&\Re({A_{kk'}+B_{kk'}})\end{array} \right), 
\end{eqnarray}
where $A_{kk'}$ and $B_{kk'}$ are Bogolubov coefficients associated with the whole trajectory \cite{NFriis:IFuentes:2012}.  In the case of a cavity initially at rest that begins to uniformly accelerate at $t=0$ the symplectic matrix, which we denote by $\mathcal{V}$, corresponds to $A_{kk'}=\alpha_{kk'}$ and $B_{kk'}=\beta_{kk'}$.  When the Bogolubov coefficients $\beta_{kk'}$ are non-vanishing there is particle creation and the cavity modes become entangled according to a comoving observer. The basic building block trajectory that corresponds to inertial-uniformly accelerated-inertial motion is implemented by the action of the symplectic matrix $S_{B}=\mathcal{V}^{-1}(h)U(\tau)\mathcal{V}(h)$. The entanglement generated between the cavity modes after a single basic building block trajectory has been analyzed in \cite{NFriis:IFuentes:2012} and found to be very small.


We are interested in trajectories that generate entangling quantum gates. 
In particular, we will show that the entanglement growth can be made cumulative by repeating trajectory 
segments that may individually contain any number of sub-segments of constant acceleration. 
We assume the accelerations of the sub-segments to be of the form $a_{i}=s_ia$ where $a$ is the largest acceleration and $s_{i}<1$. 
Assuming that $h=aL\ll1$, the Bogolubov coefficients can be expanded to first order in $h$ as 
$A_{kk'}=G_k\delta_{kk'}+A_{kk'}^{(1)}$ and $B_{kk'}=B_{kk'}^{(1)}$, where the superscript $(1)$ denotes a quantity that is of first order in~$h$, 
$G_{k}=e^{i\omega_{k}T}$ are the phases acquired by the state during segments of free evolution, 
and $T$ denotes the \textit{total\/} proper time of the segment. 
Taking the cavity to be initially in the vacuum state, 
we find that the reduced state of modes $k$ and $k'$ after an $N$-segment trajectory 
is $\sigma_N=(S_{kk'}^N)^{Tp} S_{kk'}^N$ where 
\begin{eqnarray*}
S_{kk'}&=&
\left( \begin{array}{cc}
s_{kk}&s_{kk'}\\
s_{k'k}&s_{k'k'}, 
\end{array}
\right).
\end{eqnarray*}

\emph{Two mode entangling gates.\thinspace---} 
We now show that the transformation 
$S^N_{kk'}$ corresponds to a two mode entangling gate known as the two mode squeezer. 
We find that 
\begin{eqnarray*}
\sigma_N=
\begin{pmatrix}
\mathds{1} & E_N^{(1)} \\
E_N^{(1)Tp} &  \mathds{1}
\end{pmatrix}
+\mathcal{O}(h^2),
\end{eqnarray*} 
where the $2\times2$ matrix $E_N^{(1)}$ is a function of the first order 
alpha and beta coefficients only. This state is pure, bipartite and symmetric~\cite{note}, 
and it has the form $\sigma_N=R^{Tp}Z^{Tp}Z R$ of a generalized squeezed state. 
The symplectic matrix is hence a two-mode squeezing gate, 
$S^N_{kk'}=ZR$, 
where the $4\times4$ rotation matrix $R=R(\psi_k)\oplus R(\psi_{k^{\prime}})$ is a local operation, 
\begin{eqnarray*}
Z(r)=
\begin{pmatrix}
\cosh r\mathds{1} & \sinh r \sigma_z \\
\sinh r \sigma_z &  \cosh r\mathds{1}
\end{pmatrix},
\end{eqnarray*} 
and $\sigma_x,\sigma_y,\sigma_z$ denote the Pauli matrices.
The squeezing parameter $r$ is proportional to~$h$, so that $|r|\ll1$. 
The entanglement produced by the gate will be quantified in the next section. 
We will in particular show that the the entanglement grows linearly in the number of repetitions in periodic motion. 

\emph{The entangling power of the quantum gates.\thinspace---} 
A lower bound on the entanglement generated by the quantum gate is can be found by calculating the smallest positive 
symplectic eigenvalue $\tilde{\nu}_{N}$ of the partial transposed state 
$\tilde{\sigma}_{N}=P\sigma_N P$ \cite{adesso2005} where $P=\text{diag}({1,1,1,-1})$~\cite{note2}. 
The symplectic eigenvalues are the eigenvalues of the matrix $i\Omega\tilde{\sigma}_{N}$, where the symplectic form $\Omega$ 
is for us given by $\Omega=-i\sigma_y\oplus\sigma_y$. 
This will bound a family of entanglement monotones based on the 
\textit{positive partial transpose criterion}~\cite{PhysRevLett-77-1413}, 
including the logarithmic negativity, all of which are are monotonic functions of~$\tilde{\nu}_{N}$. 

When the commutator $[S_{kk'}^{Tp},S_{kk'}]$ vanishes, the partial transposed state after $N$ segment repetitions is equal to the $N$th power of the partial transposed state after a single repetition of the segment, i.e. $\tilde{\sigma}_{N}=\tilde{\sigma}_{1}^{N}$.  
This implies that the first order correction $\tilde{\nu}^{(1)}_{N}$ to the symplectic eigenvalue grows linearly, i.e.\ $\tilde{\nu}^{(1)}_{N}=N\tilde{\nu}^{(1)}_{1}$.
Any entanglement measure $E(\tilde{\nu}_{N})$ that is a function of the symplectic eigenvalue $\tilde{\nu}_{N}$ 
then satisfies 
$E(\tilde{\nu}_{N})\sim\tilde{\nu}_{N}^{(1)}\sim N\tilde{\nu}_{1}^{(1)}\sim NE(\tilde{\nu}_{1})$, 
and therefore $[S_{kk'}^{Tp},S_{kk'}]=0$ is a \textit{resonance condition}. 
At the resonance, the logarithmic negativity is given by $E_{\mathcal{N}}=N\tilde{\nu}_{1}$, where 
$\tilde{\nu}^{(1)}_{N}\sim B_{kk'}^{(1)}$ and $B^{(1)}_{kk'}$ is the first order correction to the beta coefficients $B_{kk'}$ of the segment~\cite{NFriis:IFuentes:2012}. 
A lower bound on the entanglement generated at after $N$ segment repetitions is hence given by the logarithmic negativity, 
$E_{\mathcal{N}}=NB_{kk'}^{(1)}$. 

The commutator for an arbitrary segment, to first order in~$h$, is 
$\bigl[S_{kk'}^{Tp},S_{kk'}\bigr]
=\left(
\begin{smallmatrix}
0 & C \\
C^{Tp} & 0
\end{smallmatrix}
\right)$, 
where $C=\Re(w) \mathds{1}-\Im (w) \sigma_x$ and $w=2[(G^{*}_{k}-G_{k'})B_{kk'}^{(1)}]$.
This commutator vanishes when  
$\left(G^{*}_{k}-G_{k'}\right)B^{(1)}_{kk'}=0$. 
Recalling that the Minkowski and Rindler frequencies coincide to first order in~$h$, 
it follows that resonances occur when $B_{kk'}^{(1)}\neq0$ and the total proper time $T$ takes the discrete values 
\begin{equation}
T_n = \frac{2n\pi}{\omega_k + \omega_{k'}} \, , 
\ \ \ 
n = 1,2,\ldots 
\ . 
\label{resonant:total:times}
\end{equation}
We emphasize that $T_n$ does not depend on the details of the travel scenario; however, the total amount of entanglement generated $E_{\mathcal{N}}=NB_{kk'}^{(1)}$ does depend on the specifics of the trajectory thorough the Bogolubov coefficient. 
We further find that the average number of excitations $\left<N_k\right>$ in a cavity mode at resonance is proportional to~$\tilde{\nu}^{(1)}_{N}$.

\begin{figure}
\includegraphics[width=\linewidth]{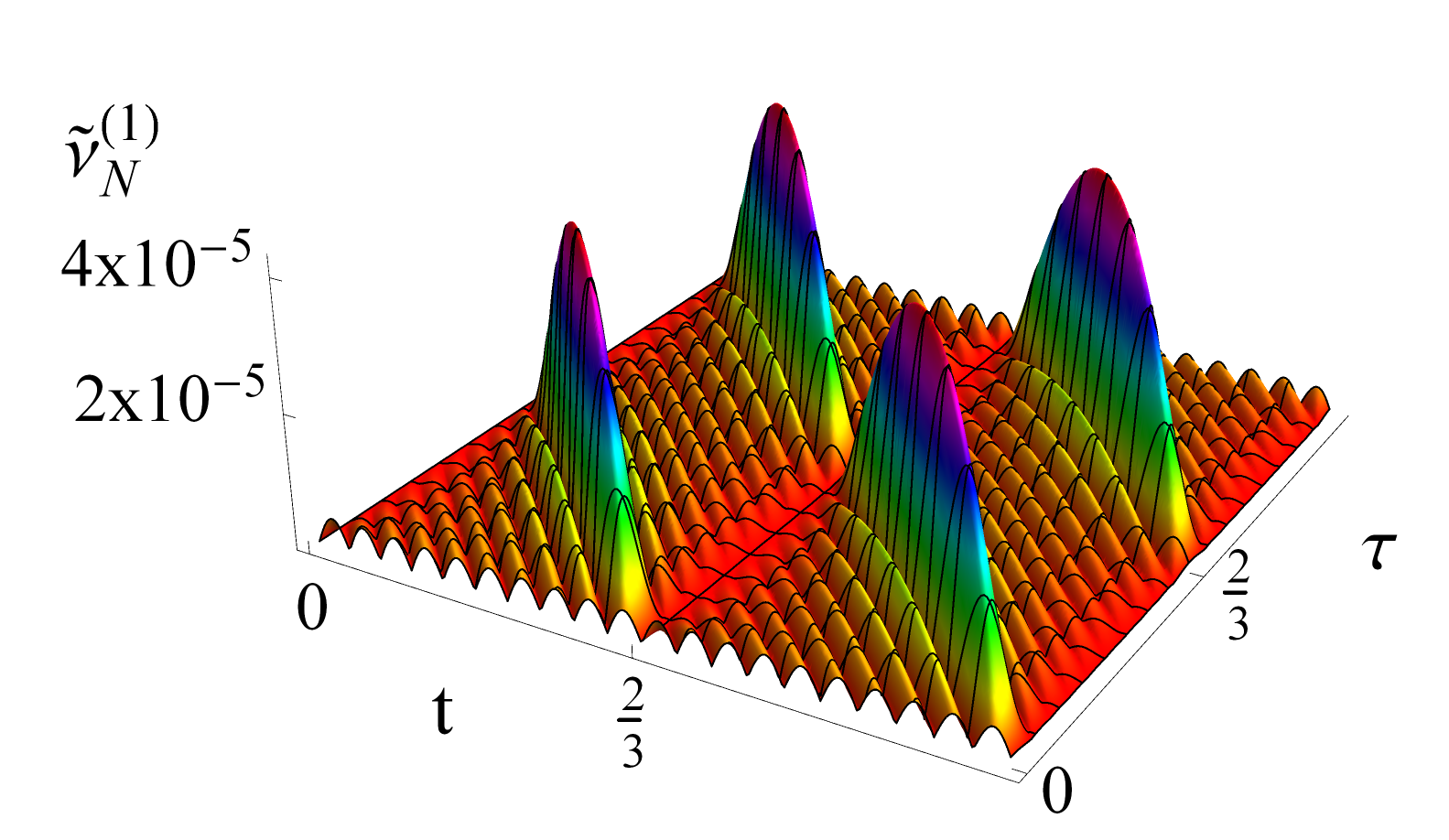}
\caption{\label{dupa} The correction to the symplectic eigenvalue $\tilde{\nu}^{(1)}_{N}$  after $N=5$ segment repetitions as a function of the proper time of acceleration~$\tau$ and the time of coasting $t$. We considered a cavity length of $L=1$, massless modes $k=1$ and $k'=2$ and accelerations $a=a'=10^{-4}$.} 
\end{figure}

\emph{Sample travel scenario.\thinspace---} 
We now specialize to a segment in which the cavity travels with proper acceleration $a=h/L$ for proper time~$\tau$, coasts for proper time~$t$, travels with proper acceleration (in either the same or opposite direction) $a'=h'/L$ for proper time $\tau$ and finally coasts for proper time~$t$. The total proper time of the segment is $T_n=2(\tau+t)$ and the symplectic transformation is
\begin{align}
S_{kk'}=&\,U_{kk'}(t)\,\mathcal{V}_{kk'}^{-1}(h')\,U_{kk'}(\tau)\,\mathcal{V}_{kk'}(h')\times\nonumber\\
&\,U_{kk'}(t)\,\mathcal{V}_{kk'}^{-1}(h)\,U_{kk'}(\tau)\,\mathcal{V}_{kk'}(h).
\end{align} 
The first order correction to the beta Bogolubov coefficient has the modulus 
\begin{equation}
|B_{kk'}^{(1)}|=c_{kk'}|1-g^{\ast}_k g^{\ast}_{k'}||1+\epsilon y g^{\ast}_k g^{\ast}_{k'}f^{\ast}_k f^{\ast}_{k'}|\,h\label{entanglement}
\end{equation}
where $g_k=\exp(i\omega_k \tau)$, $f_k=\exp(i\omega_k t)$, 
$a'=\epsilon ya$ ($y>0$), 
and $\epsilon=1$ (respectively $\epsilon=-1$) when the two accelerations have the same (opposite) direction. 
Specializing to a massless field, we have $c_{kk'}=\frac{\sqrt{kk'}(1-(-1)^{k-k'})}{\pi^2(k+k')^3}$.  
At the resonance~(\ref{resonant:total:times}), the 
logarithmic negativity is given by 
\begin{equation}
E_{\mathcal{N}}=N\,c_{kk'}|(1-(-1)^ne^{i(\omega_k+\omega_k^{\prime})t})(1+ (-1)^n\epsilon y)|\,h.\label{final:general:casimir:beta}
\end{equation}
Note that $E_{\mathcal{N}}$ vanishes when $n$ is even and the time of coasting is $t=2\pi m/(\omega_k+\omega_k^{\prime})$ and when $n$ is odd and $t=(2m+1)\pi/(\omega_k+\omega_k^{\prime})$ with $m\in\mathbb{N}$. 
Taking the accelerations to have equal magnitude ($y=1$), 
a maximal amount of entanglement is generated
for accelerations in the same direction when 
$n$ is even and $t=\pi (2m+1)/(\omega_k+\omega_k^{\prime})$, and for accelerations in opposite directions when $n$ is odd and  $t=2\pi m/(\omega_k+\omega_k^{\prime})$. 
These maxima are evident in Fig.~\ref{dupa} where we plot $\tilde{\nu}^{(1)}_{N}$ after $N=5$ segment repetitions as a function of the proper time of acceleration $\tau$ and the time of coasting~$t$.

Interestingly, the special case of sinusoidal motion corresponds to the standard dynamical Casimir setting where the cavity oscillates periodically as a whole. A~resonant enhancement of particle creation occurs in the dynamical Casimir effect \cite{dodonov1990} which was recently demonstrated in the laboratory in a superconducting circuit consisting of a coplanar transmission line with a tunable electrical length which produces an effective moving boundary \cite{wilsonnature}. In this setup it is possible to introduce a second boundary condition which, together with the first one, is modulated in such way that the system resembles a moving cavity of constant length from the perspective of a comoving observer. Furthermore, the onset of sudden accelerations has already been achieved in this system. Therefore, this setup is suitable to implement relativistic quantum gates experimentally.  In addition, in realistic scenarios cavity losses play an important role \cite{raimond2001}. We are currently investigating the loss of entanglement due to decoherence and the effects on quantum gates in the relativistic scenario considered here.  We anticipate that the results will be similar to the non-relativistic case. Decoherence will degrade entanglement and quantum gates will be therefore imperfect. 

Via the equivalence principle our results suggest that changes of the gravitational field can produce entanglement and quantum gates. For example, consider a small cavity containing a bosonic field in its vacuum state freely falling in the presence of a gravitational field \cite{Kothawala:2011fm}. Entanglement between the modes is generated by suddenly holding the cavity at a fixed position against the action of the gravitational field. If the cavity's position changes periodically or the gravitational field fluctuates, the entanglement can be enhanced. Quantifying entanglement in situations where motion or gravitation have a significant role can also provide guidance for theories about the microscopic structure of spacetime, via the Hawking-Unruh effect and its connections to thermodynamics and statistical mechanics~\cite{Hawking-BH,unruh76}.

\emph{Discussion.\thinspace---} 
We have introduced a scheme for implementing quantum processing gates in relativistic quantum field theory.  The relativistic non-uniform motion of a cavity is employed to generate paradigmatic two-mode quantum gates which produce observable amounts of entanglement. The gates can be implemented experimentally thanks to a recent breakthrough in superconducting circuits where the relativistic motion of a boundary condition was demonstrated recently~\cite{wilsonnature}.
Finding ways to create significant amounts of entanglement in relativistic settings is of great interest since entanglement is necessary for quantum communications and information processing  \cite{Schutzhold2008}. Recent studies in relativistic quantum information show that small amounts of mode entanglement are created when a cavity undergoes non-uniform motion. \cite{BruschiAlpha,friis2011}. We show that particle creation and bipartite mode entanglement can be linearly enhanced by repeating periodically any trajectory. 

\emph{Note added.} ---
After completion of this work, the generation of \emph{single\/} qubit rotations by relativistic motion was analysed in~\cite{MartinMartinez:2012uk}.

We thank Gerardo Adesso, Nicolai Friis and Davide Girolami for interesting discussions and useful comments. I.~F. and A.~D. thank EPSRC  [CAF Grant EP/G00496X/2] for financial support. A.~D. was supported in part by the Polish Ministry of Education and is funded by the National Science Center, Sonata BIS grant 2012/07/E/ST2/01402. J.~L. was supported in part by STFC~(UK).


\begin{thebibliography}{99}
\bibitem{note2}
The entanglement of a bipartite system in a pure state is quantified by the von Neumann entropy; however, this measure is not suitable in our perturbative regime because the first order contribution in the expansion of the entropy is of the form $\mathcal{E}_{VN}\sim h\ln h$ which cannot appear as a term in a power series expansion.  One can show that the symplectic eigenvalue $\tilde{\nu}_{N}$ has a well defined power series expansion in $h$.\
\end{thebibliography}

\end{document}